\documentclass[english,aps,prl,twocolumn,superscriptaddress]{revtex4}
\usepackage{epsfig}
\usepackage{graphicx}
\usepackage{amsmath}
\usepackage{xcolor}
\usepackage[normalem]{ulem}


\usepackage{babel}

\begin{document}

\title{Neutron skins as laboratory constraints on properties of neutron stars and on what we can learn from heavy ion fragmentation reactions}

\author{C.A. Bertulani}
\affiliation{Department of Physics and Astronomy, Texas A\&M University-Commerce, Commerce, TX 75429-3011, USA}
\affiliation{Institut f\"ur Kernphysik, Technische Universit\"at Darmstadt, D-64289 Darmstadt, Germany}

\author{J. Valencia}
\affiliation{Department of Physics and Astronomy, Texas A\&M University-Commerce, Commerce, TX 75429-3011, USA}

\begin{abstract}
The equation of state (EOS) of infinite nuclear matter with a small proton or neutron fraction is a crucial input to determine  the properties of neutron stars and compare model predictions to astronomical observations. The so-called ``symmetry energy'' is the part of the EOS accounting for the difference in the number of neutrons and protons. Numerous experiments have been devised to assess the symmetry energy and constrain its functional dependence with the nucleon density. Further constraints follow from a stellar modeling using the EOS to reproduce astronomical observations such as neutron star masses and radii. The recent detection of gravitational waves emitted from neutron star mergers and the nucleosynthesis ensuing from these events caused a surge of interest for such studies.  Several types of nuclear reactions have been proposed to study the symmetry energy part of the EOS.
Some of them consist in determining the neutron skin in nuclei and exploit its correlation  with the slope parameter of the symmetry energy.  In this article we explore a particular set of reactions using high energy ($E_{lab} \sim 1$ GeV/nucleon) neutron-rich projectiles. We explore measurements of all reaction fragments (a) in the same isotopic chain, i.e., only by removal of neutrons, (b) in all charge-changing channels, and (c) total interaction cross sections. Using Hartree-Fock-Bogoliubov (HBF) predictions for neutron and proton densities with Skyrme interactions, we explore the sensitivity of these cross sections with the neutron skin in nuclei. 
\end{abstract}

\date{\today}

\maketitle

\section{Introduction}
Neutron and proton distributions within a nucleus are expected to have slightly different radii. Evidences for this were accumulated from electron scattering off nuclei which yields information on their charge distribution \cite{deV87} and from hadronic probes, e.g. proton, pion or antiproton scattering \cite{Jast04}, which  are sensitive to the total nucleon distribution. The nucleus tends to builds up a small neutron skin as its number of neutrons $N$ increases.   This neutron skin in a nucleus is often quantified in terms of the  difference between its neutron and proton distribution rms radius, i.e.,
\begin{equation}
\Delta r_{np} = \left<r_{n}^{2}\right>^{1/2}- \left<r_{p}^{2}\right>^{1/2}.
\end{equation}
Theoretical values for this quantity are deduced from microscopic models for the nucleus such  as Hartree-Fock-Bogoliubov theories using Skyrme or Gogny interactions \cite{RS04}.  The neutron skin size predictions range from zero for light stable nuclei, with $N\approx Z$, up to about 0.3 fm for heavy nuclei with large neutron excess \cite{Dut12,Bend03,Ston07,Sell14,Roc11,Roc15,FPH17}. However, using all experimental techniques presently available, it has not yet been possible to determine the neutron skin accurately enough in order to discern the best microscopic theories for proton and neutron distributions \cite{Aum17}.

The neutron excess in a nucleus leads to a corresponding neutron pressure larger than that for the proton component. Such nucleon pressure is closely related to the energy per nucleon, $\epsilon=E/A$, which is a function of the nuclear density and its nucleon asymmetry. For a given set of neutron $N$, proton $Z$, and mass $A=N+Z$ numbers,  the energy per nucleon around $N=Z$ is given by
 \begin{equation}
\epsilon_{A}(\rho,\delta) = \epsilon_{A} (\rho, 0) + S_{A}(\rho) \delta^{2}+\cdots,\label{epsilon}
\end{equation}
where  $\delta = (N-Z)/A$ is the asymmetry coefficient and $S_{A}$ is known as the symmetry energy. In neutron stars, an ``infinite``' nuclear matter environment, with $A\approx 10^{57} \rightarrow \infty$, one usually drops the index $A$ for the physical quantities above and the asymmetry coefficient is described in terms of the neutron, proton and total densities, $\delta = (\rho_{n}-\rho_{p})/\rho$. The density dependence of $S$ around the nuclear matter saturation density $\rho_0\simeq 0.16$ fm$^{-3}$ is obtained from a Taylor expansion,
\begin{equation}
S(\rho) =J+{\L\over 3} {{\rho-\rho_{0}}\over \rho_{0}}+\cdots, \label{symme}
\end{equation}
where $J=S(\rho_{0})$ is the  bulk symmetry energy and $L=3\rho_{0}dS(\rho)/d\rho|_{\rho_{0}}$ determines its slope. At the saturation density, $\rho_{0}=0.16$ fm$^{-3}$, the binding energy per nucleon is  $\epsilon(\rho_0,0)\simeq -16$ MeV. There is a strong experimental evidence that the value of $J\approx 30$ MeV is compatible with theoretical predictions.

The pressure in homogeneous nuclear matter, determining the equation of state (EOS), is given by 
$p(\rho,\delta)= \rho^{2}{d\epsilon(\rho,\delta) / d\rho}$.
From Eqs. \eqref{epsilon} and \eqref{symme}, the EOS is therefore strongly dependent on the symmetry energy, $S$. In fact, for pure neutron matter, $\delta =1$, and for $\rho$ close to $\rho_0$ one has $p = L\rho_0/3$, emphasizing the importance of the slope parameter $L$. 
The quantity $L$ is poorly determined experimentally with its value varying within $0$ and 150 MeV, theoretically \cite{Roc11,Roc15}. For finite nuclei,  the competition between the two terms in Equation \eqref{symme}  influences the  thickness of the surface region where the neutron contribution is dominant for $N>Z$. Therefore, we expect that studies of neutron skin in nuclei will reveal the details of the EOS needed to perform calculations for the structure of neutron stars, with $\delta =1$. The EOS of nuclear matter is also needed to describe the  explosion mechanism of core-collapse supernovae \cite{Gle97,Web05,Hae07,LP04,LP12,BP12,Heb13}

There has been several nuclear physics experiments and astronomical observations aimed at studying the EOS of nuclear matter and the role of the symmetry energy. As a subset of these, the measurement of neutron skins in nuclei has also attracted large experimental interest \cite{Aum17,Tam11,Ros13,Shub18}. The width of the neutron skin can also be studied with different experimental techniques, although an accurate measurement is still lacking. Electromagnetic probes of the nucleus can infer their radial charge distribution rather well, but the determination of its matter distribution is still difficult. Recently, experimental efforts were proposed to deduce the neutron skin from accurate measurements of fragmentation reactions using inverse kinematic collisions and light targets such as carbon and proton targets \cite{Aum17}. Inverse kinematics with radioactive beams allows one to use nuclear projectiles with varying neutron number along an isotopic chain. This technique is crucial, because many of the nuclei along the chain are short-lived requiring their use as projectiles. For example, one can measure charge-changing cross sections $\sigma_{{\Delta Z}}$, i.e., the total cross sections for the production of fragments with one or multiple protons removed from the projectile. Another possibility is to  measure neutron-changing cross sections, $\sigma_{\Delta N}$, accounting for all fragments with at least one neutron removed \cite{Aum17}. 

At high projectile energies, it is often considered a good assumption that the fragmentation reaction occurs in two steps. In the first step primary fragments are produced by scrapping nucleons off the projectile,  and a second step occurs when the energy deposited in the fragments leads them to undergo a nuclear decay with the emission of $\gamma$, $\alpha$-particle, nucleon evaporation, etc. This is a purely theoretical assumption since there is no accurate experimental procedure to separate the two steps. Moreover, the second step is also theoretically difficult to calculate accurately. It is frequently modeled using statistic models, such as the Hauser-Feshbach theory \cite{HF52,Fesh93}. This theory needs several input parameters such as level-densities and barrier transmission probabilities which are still under intense scrutiny.  

The first step in the reaction, as mentioned above, is easier to model with help of methods like the Glauber model for nuclear collisions \cite{Gla59,Fesh93,BD04,Mill07}. The Glauber model is not free of uncertainties either, but it contains a much smaller number of assumptions and has been used to describe with success an enormous number of experiments on high energy hadronic reactions. That is exactly why  a measurement of $\sigma_{\Delta N}$ or $\sigma_{\Delta Z}$ is advantageous: One avoids the need for a theoretical description of all possible nucleon evaporation decays  in the reaction second step. Another possibility is to measure the total interaction cross section $\sigma_{I} = \sigma_{\Delta N} + \sigma_{\Delta_{Z}}$ defined here as the cross section for the removal of at least one nucleon, irrespective if they are protons or neutrons.

The purpose of this work is to study the sensitivity of   $\sigma_{I}$, $ \sigma_{\Delta N}$, and $\sigma_{\Delta {Z}}$ on the neutron skin of nuclei and consequently on the most uncertain part of the symmetry energy, namely its slope parameter defined in Eq. \eqref{symme}. In the next sections we will present a summary of the assumptions we employ to obtain nucleon removal cross sections and the contributions from several different mechanisms. After that we present our numerical results followed by our conclusions 

\section{Theoretical Modeling of Neutron- and Charge-Changing Reactions}

The Glauber model has been widely adopted in calculations of numerous processes in high energy nuclear collisions \cite{Gla59,Fesh93,BD04,Mill07}. Here we will employ it to calculate the cross section to produce a primary fragment with charge and neutron number $(Z_{F},N_{F})$ in the collision of  a projectile nucleus with charge and neutron number $(Z_{P},N_{P})$ with a nuclear target. The cross sections are deduced from \cite{BD04,Mill07,Aum17}
\begin{eqnarray}
\sigma(Z_{F},N_{F})&=&
\left(
\begin{array}{c}Z_{P} \\ Z_{F}
\end{array}
\right)
\left(
\begin{array}{c}N_{P} \\ N_{F}
\end{array}
\right)
\int d^2 b \left[ 1-P_p(b)\right]^{Z_{P}-Z_{F}} \nonumber \\
&\times& P_p^{Z_{F}}(b)\left[ 1-P_n(b)\right]^{N_{P}-N_{F}}P_n^{N_{F}}(b), \label{sigma}
\end{eqnarray} 
where $b$ is the impact parameter in the collision. The binomial coefficients take into account  all possible ways that $Z_{F}$ protons can be removed from the $Z_{P}$ initial protons of the projectile. A similar counting is made for the  neutrons. $P_p$ ($P_n$) are the probabilities for the survival of a single proton (neutron) of the projectile and the factors containing $(1-P)$ account for the removal probability of the other protons (neutrons).  $P_p$  and $P_n$ are given by   \cite{BD04,Mill07}
\begin{eqnarray}
P_p(b)&=&\int dzd^2s \rho_p^P({\bf s},z) \exp\left[ -\sigma_{pp} Z_T\int d^2s \rho_p^T({\bf b-s},z) \right. \nonumber \\
&-&\left. \sigma_{pn} N_T\int d^2s \rho_n^T({\bf b-s},z) \right],  \label{ppb}
\end{eqnarray} 
where the charge and neutron number of the target is denoted by $(Z_{T},N_{T})$, and $\rho_p$ ($\rho_n$) is the proton (neutron) density of projectile and target,  normalized to unity. $\sigma_{np}$ and $\sigma_{pp}$ are the neutron-proton and proton-proton (without Coulomb) total cross sections. At high energies, it is assumed that medium effects are small and the nucleon-nucleon cross sections are taken from a fit of experimental data of free nucleon scattering at energies in the range $E_{lab} = 500$ to $5000$ MeV  \cite{BC10}.  Similarly, for $P_{n}$ we use
\begin{eqnarray}
P_n(b)&=&\int dzd^2s \rho_n^P({\bf s},z) \exp\left[ -\sigma_{pp} N_T\int d^2s \rho_n^T({\bf b-s},z) \right. \nonumber \\
&-&\left. \sigma_{pn} Z_T\int d^2s \rho_p^T({\bf b-s},z) \right].  \label{pnb}
\end{eqnarray} 

As proposed in Ref. \cite{Aum17}, the neutron skin, and its evolution with increasing number of neutrons of the projectiles, can be extracted by the application of the theory described above. The idea is to study measurements of  $\sigma_{\Delta N}$,  the cross section to produce all isotopes of the projectile by removing at least one of its neutrons. It is obtained by setting $Z_{F}=Z_{P}$ in Eq. \eqref{sigma} and adding from $N_{F}=1$ up to $N_{P}$. Compact equations can be obtained by using sums over the binomial coefficients \cite{BD04}, e.g., 
\begin{eqnarray}
\sigma_{\Delta N}=
\int d^2 b \left[P_p(b)\right]^{Z_{P}}\left[1- \left[1-P_n(b)\right]^{N_{P}}\right] . \label{sigmaDN}
\end{eqnarray} 
Experimentally, one intends to exploit an isotopic chain, e.g., tin isotopes, and compare the measurements with calculations employing neutron and proton density distributions based on well established microscopic theories \cite{Aum17}. To date, and for most of the heavy nuclear isotopes, such theories are often based on mean field methods, e.g., Hartree-Fock or relativistic mean-field theories \cite{RS04}. 

Similarly to neutron-changing, one could measure charge-changing, $\sigma_{\Delta Z}$, or total interaction cross sections, $\sigma_{I}$, to assess information on the neutron skin of the projectiles. The charge-changing cross sections, $\sigma_{\Delta Z}$, need to include the measurement of all elements produced out of the projectile, plus all their corresponding isotopes. The cross sections are much larger because of the many more  possibilities involved, as deduced from Eq. \eqref{sigma}. The interaction cross section is obtained by summing all possibilities that at least one nucleon is removed, $\sigma_{I}=\sum_{Z_{F},N_{F}}\sigma(Z_{F},N_{F})$. 

In the literature, the optical limit of the Glauber theory is often used. In this limit one assumes that $P(b) \ll 1$ so that one can replace $1-P(b) \approx \exp[-P(b)]$ \cite{BD04,Hor07}. Sometimes $P_{n}$ and $P_{p}$ are also assumed to be the same. Evidently, all these approximations are not appropriate to study the effects of the neutron skin. We will therefore use expression \eqref{sigma} in our analysis. And, since $\sigma_{\Delta Z}= \sigma_{I}- \sigma_{\Delta N}$,  conclusions for $\sigma_{\Delta Z}$ can be easily drawn from the knowledge of the first two cross sections. Similar studies to access information on the neutron skin in nuclei using the Glauber theory for high energy scattering have been published in, e.g., Refs. \cite{Tran16,Kan16,Yam11,Suz16}.

It is worthwhile to mention that the fragmentation process described above neglects the possibility that the projectile remains the same nucleus after a primary interaction with the target, but is excited to a collective giant resonance. Giant resonances lie above the nucleon emission threshold. As a result of the Coulomb barrier the nucleus will generally emit neutrons, and often just one neutron. Their excitation will thus contribute to the neutron-changing cross section, $\sigma_{\Delta N}$, and to the interaction cross section, $\sigma_{I}$.  The excitation of  giant resonances will be minimized if one uses light targets such as  carbon or proton. Experimentally, one can also try to disentangle this process from the nucleon stripping process described above by comparing the energy dependence of the cross sections and/or using different targets.

To estimate the cross sections for excitation of giant resonances (GR) we work within the first-order perturbation theory. We also assume that Coulomb and nuclear interference is small so that we can separate the Coulomb and nuclear excitation cross sections. As we will show later this is not so relevant as the Coulomb cross sections are much smaller than the multi-nucleon stripping cross sections defined via Eq. \eqref{sigma}. It will also be smaller than the nuclear induced excitation of GRs for light targets. The cross sections for Coulomb excitation are largest for electric dipole (E1) excitations and in particular for the isovector giant dipole resonance (GDR). It leads overwhelmingly to neutron decay and can be calculated as \cite{BB88,ABS95}
\begin{equation}
\sigma_{C}^{-n}=\int {dE\over E} n_{E1}(E) \sigma_{\gamma}^{GDR}(E),
\end{equation}
where the equivalent photon number is given by
\begin{eqnarray}
n_{E1}(E) &=&{2Z_{T}^2\alpha \over \pi}\left( {\omega c\over \gamma v^2}\right)^{2}\int db\ b \nonumber \\
&\times& \left[K_1^2(x) + {1\over \gamma^{2}}K_{0}^{2}(x) \right] \Lambda(b), \label{ne1}
\end{eqnarray}
with  $v$ being the projectile velocity, $\gamma = (1-v^2/c^2)^{-1/2}$ is the Lorentz contraction factor, $\alpha$ is the fine-structure constant and $K_{n}$ is the modified Bessel function of nth-kind, as a function of $x=\omega b/ \gamma v$, where the excitation energy is $E=\hbar \omega$. The photo-nuclear cross sections $\sigma_{\gamma}^{GDR}$ are calculated by assuming a Lorentzian shape 
\begin{equation}
\sigma_{\gamma}^{GDR}(E)=\sigma_{0}{E^{2}\Gamma^{2}\over (E^{2}-E_{GDR}^{2})^{2}+E^{2}\Gamma^{2}},
\end{equation}
where $E_{GDR} = 31.2A_{P}^{-1/3} + 20.6A_{P}^{-1/6}$ reproduces the mass dependence of the centroid of the experimentally measured GDR. It is a mixture of the excitation energy mass dependence predicted by Goldhaber-Teller and Steinwedel-Jensen macroscopic models  \cite{GT48,SJ50}. The parameter $\sigma_0$ is chosen to reproduce the Thomas-Reiche-Kuhn (TRK) sum rule 
\begin{equation}
\int dE \sigma_{\gamma}^{GDR}(E)=60{N_{P}Z_{P}\over A_{P}}\ {\rm MeV.mb}, \label{gdr}
\end{equation}
which is obtained from a nearly model independent account of the full nuclear response to a dipole operator \cite{EG87}. 

The width $\Gamma$ of the GDR is a more complicated issue. It is strongly dependent on the shell structure of the nuclei. The experimental systematics yield values for the width ranging from 4 to 5 MeV for closed-shell nuclei up to 8 MeV for nuclei situated between closed shells. Since most nuclei considered in this article are not amenable to experimental investigations using photo nuclear reactions, a possibility is to adopt a microscopic theoretical model such as the random phase approximation (RPA)  \cite{BP99}. Instead, we will use here a simple phenomenological parametrization of the GDR width in the form, $\Gamma_{GDR} = 2.51 \times 10^{-2} E_{GDR}^{1.91}$ MeV, with $E_{GDR}$ in units of MeV \cite{Pluk18}.   

The profile function $\Lambda$ in Eq. \eqref{ne1} is given by
 \begin{equation}
\Lambda(b)=\exp\left[ -\sigma_{NN}(E_{{NN}}) \int dz \int d^{3}r \rho_{P} (r) \rho_{T} (|{\bf R}-{\bf r} |)  \right] ,  \label{lambda}
\end{equation}
where ${\bf R}=({\bf b},z)$, with $z$ being the coordinate along the projectile velocity and ${\bf b}$ being the coordinate perpendicular to it. $\rho_{i}$ ($i=P,T$) are the total nucleon densities of the projectile and target, assumed to be spherical. $\sigma_{NN}$ is the isospin averaged nucleon-nucleon cross section,  parametrized as in Ref. \cite{BC10}.  

The contribution of the nuclear interaction to the excitation of giant resonances followed by neutron emission can be obtained employing a first-order (eikonal-DWBA) reduction of the coupled-channels treatment discussed in Ref. \cite{BCG03}. At high energies only the nuclear excitations of the IVGDR, $L=1$, and of the isoscalar giant quadrupole resonance (ISGQR), $L=2$,  are of relevance. The cross sections are obtained by integrating the inelastic scattering amplitude using eikonal scattering waves \cite{BD04}
 \begin{equation}
f_{L}(\theta) = {ik\over 2\pi \hbar v} \int e^{i{\bf q}\cdot {\bf R}+i\chi(b)}U_{L}(R)d^{2}bdz,
\end{equation} 
where $\chi$ is the eikonal phase, $k$ the projectile momentum and ${\bf q}$ is the momentum transfer in the reaction. The potential for the excitation of  the $L$-type resonance is often based on the deformed potential model \cite{Sat87},
 \begin{equation}
U_{L}(R)=-{\beta_{L} \over \sqrt{2L+1}}Y_{L0}(\hat {\bf R}){dU_{opt}\over dR}, \label{ul}
\end{equation} 
where $\beta_{L}$ is the deformation parameter and $U_{opt}$ is the optical potential. The deformation parameters for the IVGDR and the ISGQR are deduced from a full exhaustion of the sum-rules for dipole and quadrupole operators. They are  \cite{Sat87}
 \begin{equation}
\beta_{1}=\left({\pi \hbar^2\over 2m_{N}}{A_{P}\over N_{P}Z_{P}E_{GDR}}\right)^{1/2}{3\Delta r_{np}\over 2R_{0}}, \label{beta1}
\end{equation} 
and
 \begin{equation}
\beta_{2}=\left({20\pi \hbar^2\over 3m_{N}}{1\over A_{P}E_{GQR}}\right)^{1/2}, \label{beta2}
\end{equation} 
where $m_{N}$ is the nucleon mass, $\Delta r_{np}$ is the neutron skin and $R_{0}$ is the mean nuclear radius. The centroid of the ISGQR is taken as $E_{ISGQR}=62/A_{P}^{1/3}$ MeV.

The differential cross section is given by $d\sigma_{L}/d\Omega = |f_{L}(\theta)|^{2}$.  Since in high-energy collisions $q \simeq k \sin\theta$, where $\theta$ is the scattering angle, the solid scattering angle is given by $d\Omega = d^{2}q/k^2$ and the integration over angles can be done trivially, yielding a $\delta$-function. The total cross section for multipolarity $L$ becomes
 \begin{equation}
\sigma_{L}=\int d^{2}b \Lambda (b) |u_{L}(b)|^{2},
\end{equation} 
where $\Lambda (b)=\exp\left\{2{\it Im}[\chi(b)]\right\}$ is the same profile function as in Eq. \eqref{lambda} and
 \begin{equation}
u_{L}={1\over \hbar v} \int U_{L}(R)dz
\end{equation} 
is a dimensionless transition potential. We have also performed calculations using the coupled-channels code DWEIKO \cite{BCG03} assuming only two states (IVGDR and ISGQR) located at their centroid energies and carrying the full sum rule strengths. The results obtained are nearly identical with those using the formulation presented above. 

One of the main sources of uncertainty in the calculation of nuclear excitation stems from the optical potential. There are no optical potentials extracted from experimental systematics to describe nucleus-nucleus scattering available for the isotopic chains and the  high energies we consider here. Therefore, we resort to the ``t$\rho\rho$'' optical potential, as described in Ref. \cite{BCG03}. It is also worth noticing that the nuclear excitation of the IVGDR is directly correlated with the neutron skin, as is explicitly shown in Eq.  \eqref{beta1}. However, as we will show later, the cross sections for the excitation of the ISGQR are much larger and this correlation is not very useful to explore in this context. It is also difficult from measurements of the neutron removal reactions at relativistic energies to identify which type of resonance contributed to the cross section. 

Another sort of reaction mechanism can contribute to the charge-changing and neutron-changing reactions, namely, charge-exchange reactions. At high projectile energies, the isospin-dependent part of the NN interaction induces a change by one or more units of charge in the projectile accompanied by the opposite sign change in the target.  Microscopically, this can be viewed as the exchange of a charged pion, or a charged $\rho$. The cross sections for this process are rather small, no more than a few millibarns and will not be considered here \cite{BD04}. 

\section{Proton and neutron densities and the EOS of nuclear matter}

Numerous theoretical methods exist to obtain nucleon density distribution in nuclei. We will only consider Skyrme interaction models together with the  Hartree-Fock-Bogoliubov (HFB) theory. For a review, see, e.g., \cite{RS04}. The Skyrme interactions are contact interactions with several terms accounting for coordinate, spin and isospin dependence. With them, rather simple numerical procedures have been developed to calculate the binding energy of nuclei and several other nuclear properties. As a byproduct, the energy density functional $E[\rho]$ is obtained. From this density dependence of the nuclear energy one can try to infer many of the properties of neutron stars \cite{Gle97}.

 We have considered a sufficiently large collection of Skyrme interactions to test the dependence of the cross sections on the neutron skin of nuclei, but, in contrast to many previous studies, there is no intent here to check if one interaction does a better job than another. To each of the interactions we added a mixed pairing interaction of the form
  \begin{equation}
v({\bf r}, {\bf r}') = v_{0}\left( 1- {1\over 2}{\rho \over \rho_{0}}\right)\delta({\bf r}-{\bf r}'),
\end{equation} 
where $\rho(r)=\rho_{n}(r) + \rho_{p}(r)$ is the isoscalar local density, the pairing strength adopted is the same for neutrons  and protons, $v_{0}=-131.6$ MeV, and the saturation density is fixed at $\rho_{0}=0.16$ fm$^{-3}$. Our calculations were performed with the code HFBTO \cite{Sto05}. The zero range character of the pairing force requires the introduction of a cutoff energy in the quasiparticle  space, and we have chosen $E_{cut} = 60$ MeV.

We have used parameters for the Skyrme interactions available on the CompOSE (Compstar) repository at
https://compose.obspm.fr/. We have included the following Skyrme interactions: SIII \cite{Bei75}, SKA and SKB \cite{Koh76},  SKM* \cite{Bar82}, SKP \cite{Dob84}, UNE0 and UNE1 \cite{Sto13}, SKMP \cite{Ben89}, SKI2, SKI3, SKI4 and SKI5 \cite{Rei95}, SLY230A \cite{Cha97}, SLY4, SLY5, SLY6, and SLY7 \cite{Cha98},  SKX \cite{Bro98}, SKO \cite{Rei99},  SK255 and SK272 \cite{Arg03}, HFB9 \cite{Gor05} and SKXS20 \cite{Dut12}. The calculations for the ENE0 and UNE1 interactions were done with the modified version of the code HFBTO code \cite{Sto13}. 

The large number of Skyrme interactions employed here also leads to a large variation of nuclear matter properties.  This is summarized for a few of these interactions in Table \ref{NMp}. The incompressibility of nuclear matter is defined as 
  \begin{equation}
K_{0}= 9 \rho_{0}^{2}\left. {\partial^{2}E/A \over \partial \rho^{2}}\right|_{\rho_{0}},
\end{equation} 
and $J$ and $L$ have been introduced in Eq. \eqref{symme}. It is clear that the least constrained EOS property is the slope of the symmetry energy, $L$. Because these interactions have been fitted to reproduce several nuclear properties, it is very hard to judge which one would be better suited to study  neutron star properties.

\begin{table}[ht]
\begin{center}
\caption{Nuclear matter properties at saturation density associated with different Skyrme interactions. All quantities in MeV. \label{NMp}}
\begin{tabular}{|c|c|c|c||c|c|c|c|c|}
\hline\hline
Skyrme & $K_{0}$ & $J$   &  $L$ & Skyrme & $K_{0}$ & $J$   &  $L$   \\ 
\hline 
SIII & 355. & 28.2 & 9.91& SLY5 & 230. & 32.0 & 48.2 \\
\hline 
SKP & 201. & 30.0 & 19.7& SKXS20 & 202. & 35.5 & 67.1\\
\hline
SKX & 271. & 31.1 & 33.2 & SKO & 223. & 31.9 & 79.1 \\
\hline
HFB9 & 231. &  30.0 & 39.9   & SKI5 & 255. & 36.6 & 129.      \\
\hline
\hline
\end{tabular}
\end{center}
\end{table}

\begin{figure}[t]
\centerline{
\includegraphics[width=0.95\columnwidth]{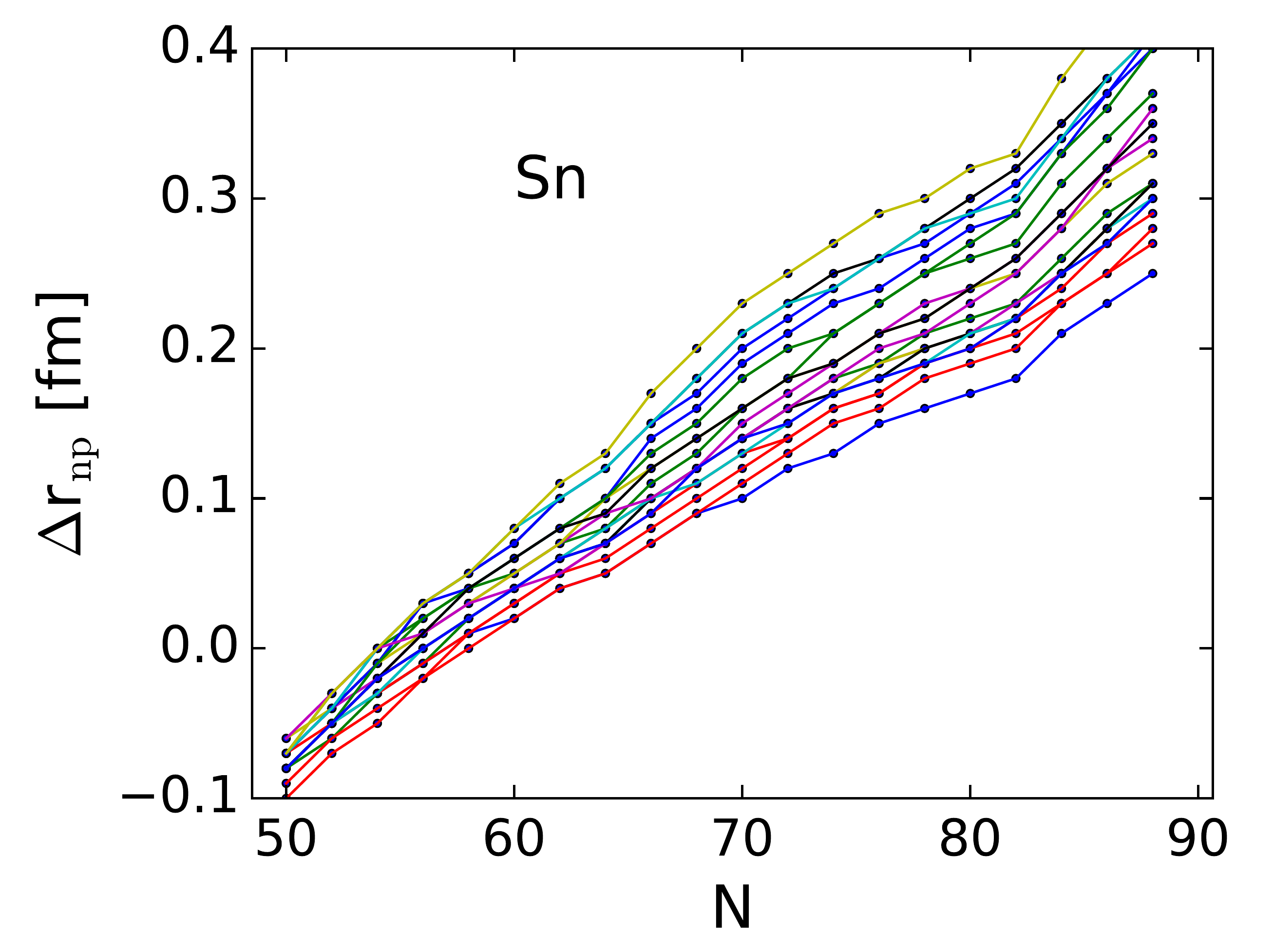}}
\caption{(color online) The points represent neutrons skin, $\Delta r_{np}$ calculated for tin isotopes with the 23 Skyrme interactions listed in the text. Each one of the lines corresponds to one of the interactions and is also a guide to the eyes.}
\label{fig1}
\end{figure}

In Figure \ref{fig1} we plot the neutron skin, $\Delta r_{np}$ calculated with the 23 Skyrme interactions listed previously. Notice that the neutron skins obtained with different Skyrme interactions tend to diverge from each other as the neutron number increases. A similar trend is observed for Ni and Pb isotopes. For the stable tin isotopes with mass $A=116, \ 118,\ 120$, the neutron skin varies in the range $0.1-0.3$ fm depending on the Skyrme model adopted. 

\begin{figure}[t]
\centerline{
\includegraphics[width=0.95\columnwidth]{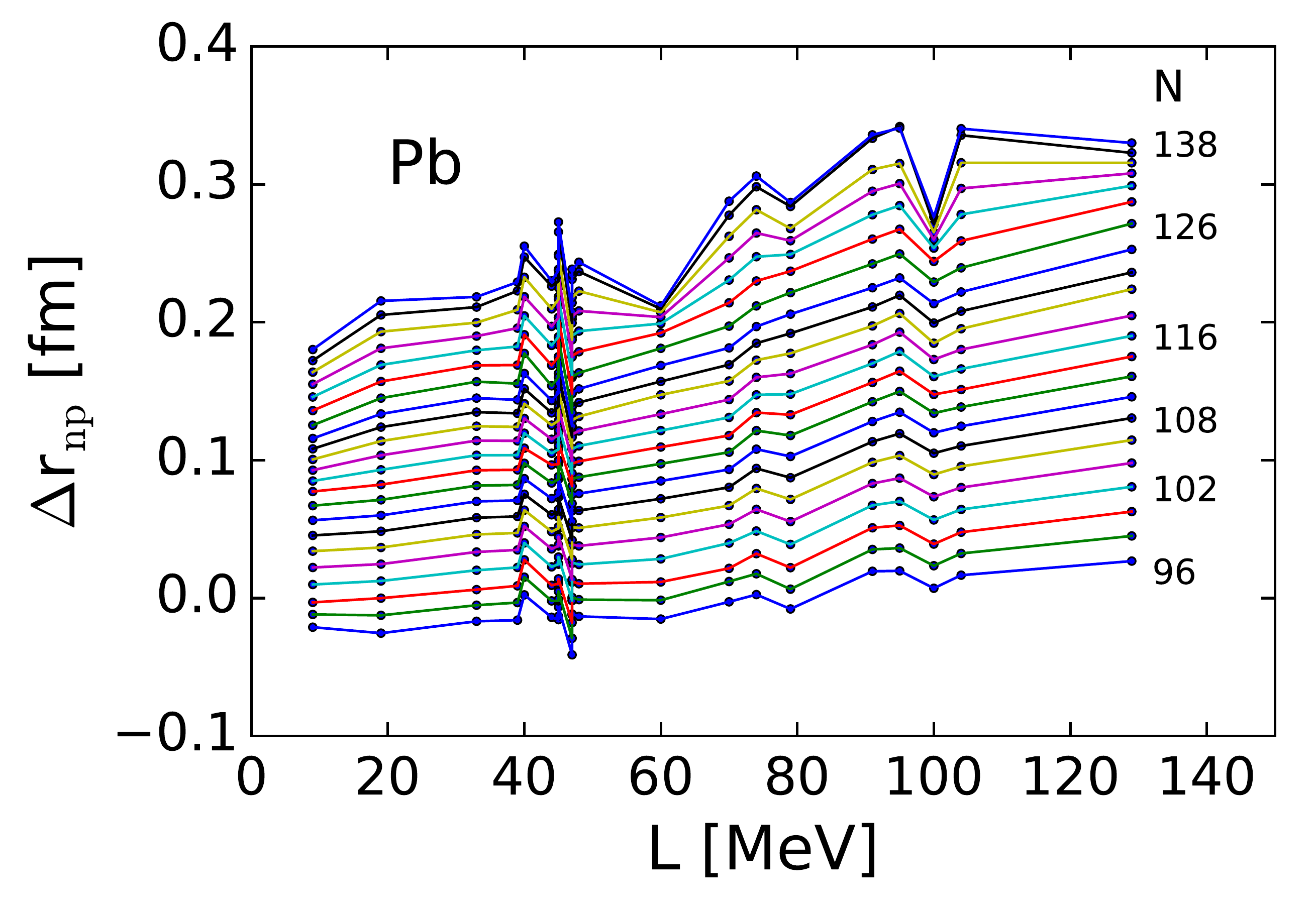}}
\caption{(color online) Neutrons skin, $\Delta r_{np}$ calculated with the Skyrme interactions listed in the text displayed as function of the value of the slope parameter $L$ predicted by each one of them. The lines are guides to the eyes. Each curve corresponds to a different lead isotope with neutron number $N$.}
\label{fig2}
\end{figure}

Despite these different isotopic dependencies, a linear correlation between $L$ and $\Delta r_{np}$ has been found for  $^{208}$Pb using  both relativistic and non-relativistic mean field models \cite{Roc11,Bro00,TB01,Fur02}. This correlation is therefore useful for planning studies of neutron skins in nuclei and extracting the value of the slope parameter $L$. This is again explored in Figure \ref{fig2}, where the neutrons skin, $\Delta r_{np}$, calculated with the Skyrme interactions listed previously are displayed as function of the value of the slope parameter $L$ predicted by each one of them. The lines are guides to the eyes. Each curve corresponds to a different value of the neutron number. One notices that, except for a few kinks, there is indeed a nearly linear relation between $L$ and $\Delta r_{np}$ even for different isotopes of lead. If the interactions predicting  $L\sim 45$ MeV values are neglected, a better linear correlation would become evident. However, we will not use this sort of argument to discriminate against any of the interactions and we will keep all listed interactions in our numerical calculations of the fragmentation cross sections.

\section{Numerical results}

\subsection{Coulomb excitation followed by neutron emission}

\begin{figure}[t]
\centerline{
\includegraphics[width=0.95\columnwidth]{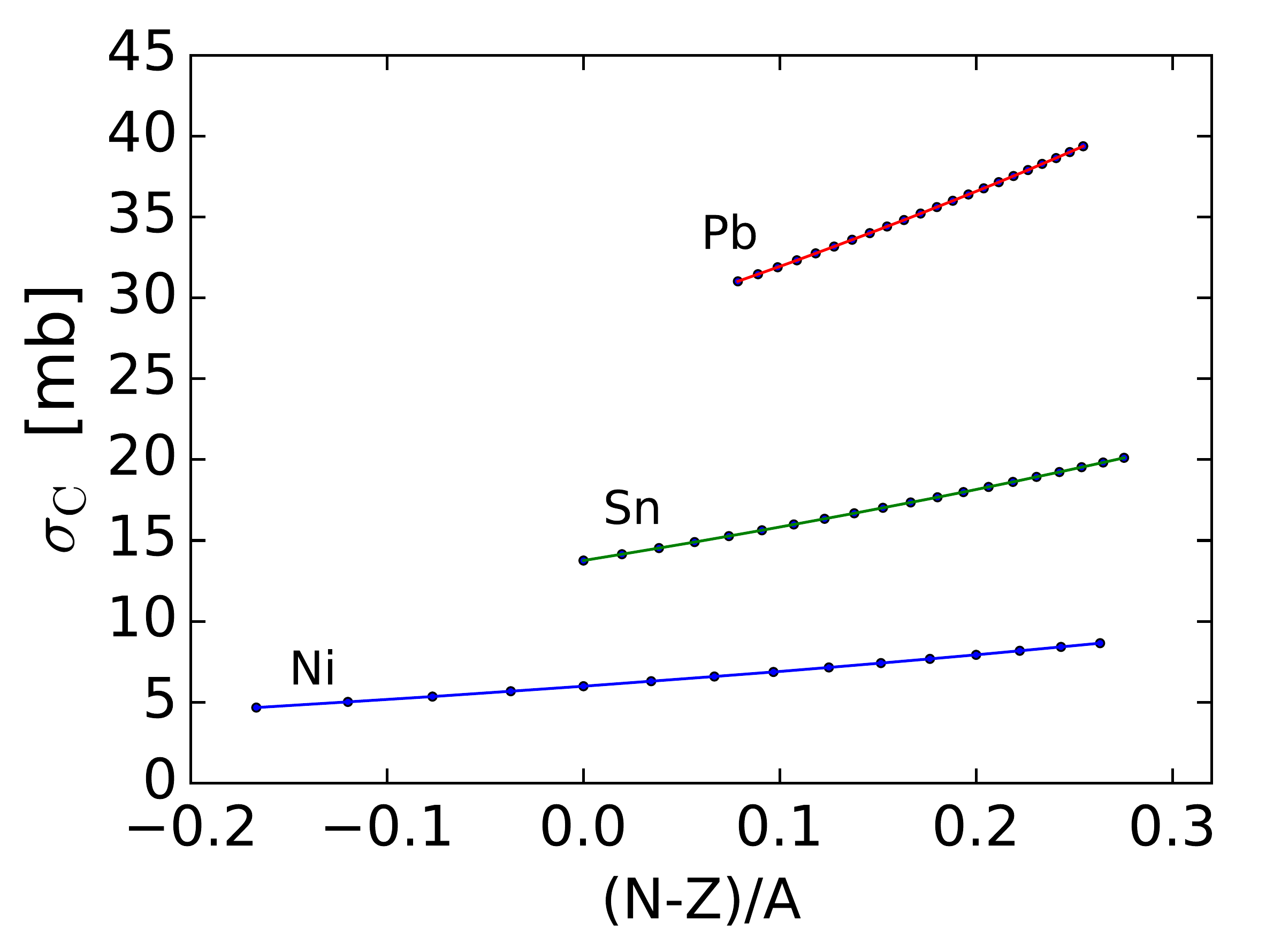}}
\caption{(color online) Cross sections in millibarns for the Coulomb excitation of the IVGDR in nickel, tin and lead projectiles incident on carbon targets at 1 GeV/nucleon, as a function of the asymmetry coefficient $\delta=(N-Z)/A$ of the projectile.}
\label{fig3}
\end{figure}

We now discuss what could be considered as ``small'' corrections to the nucleon removal cross sections  $\sigma_{\Delta N}$ and $\sigma_{I}$. We start with the Coulomb excitation of giant resonances followed by neutron emission. In  Figure \ref{fig3} we show our results for the cross sections for the excitation of the IVGDR in nickel, tin and lead projectiles incident on carbon targets at 1 GeV/nucleon, as a function of the asymmetry coefficient $\delta=(N-Z)/A$ of the projectile. The Coulomb cross section has little dependence on the neutron skin, except for the cutoff at small impact parameters through the function $\Lambda(b)$ appearing in Eq. \eqref{lambda}, but the cross section depends strongly on the asymmetry coefficient $\delta$, mainly because of the photo nuclear cross section isotopic dependence, a legacy of Eq. \eqref{gdr}. Also important is the mass dependence of the centroid energy of the resonance because the virtual photon numbers $n_{E1}$ vary strongly with the excitation energy.

As seen from Fig. \ref{fig3} the cross sections increase almost linearly with $(N-Z)/A$. For nickel and tin isotopes they are very small and can be neglected compared to the neutron changing and interaction cross sections, as we will show later. However, for lead projectiles the cross sections are not so small if one uses carbon targets. Since the Coulomb cross sections are nearly proportional to the square of the charge of the target, for proton targets the cross sections are smaller than for carbon targets by about a factor of $30-40$ and are therefore negligible. Hence, by comparing experimental results with carbon and proton targets one can easily eliminate the Coulomb cross sections contributions to the fragmentation cross sections. More elaborate theories for Coulomb excitation followed by neutron evaporation can also be employed if one wants to use a light target such as carbon. The comparison with experimental data has been shown to be nearly perfect \cite{ABS95}. There exists a large variation of the Coulomb cross sections with bombarding energy, as shown in Ref. \cite{BB88} which also helps disentangle their contribution from the other processes.

\subsection{Nuclear excitation followed by neutron emission}

\begin{figure}[t]
\centerline{
\includegraphics[width=0.95\columnwidth]{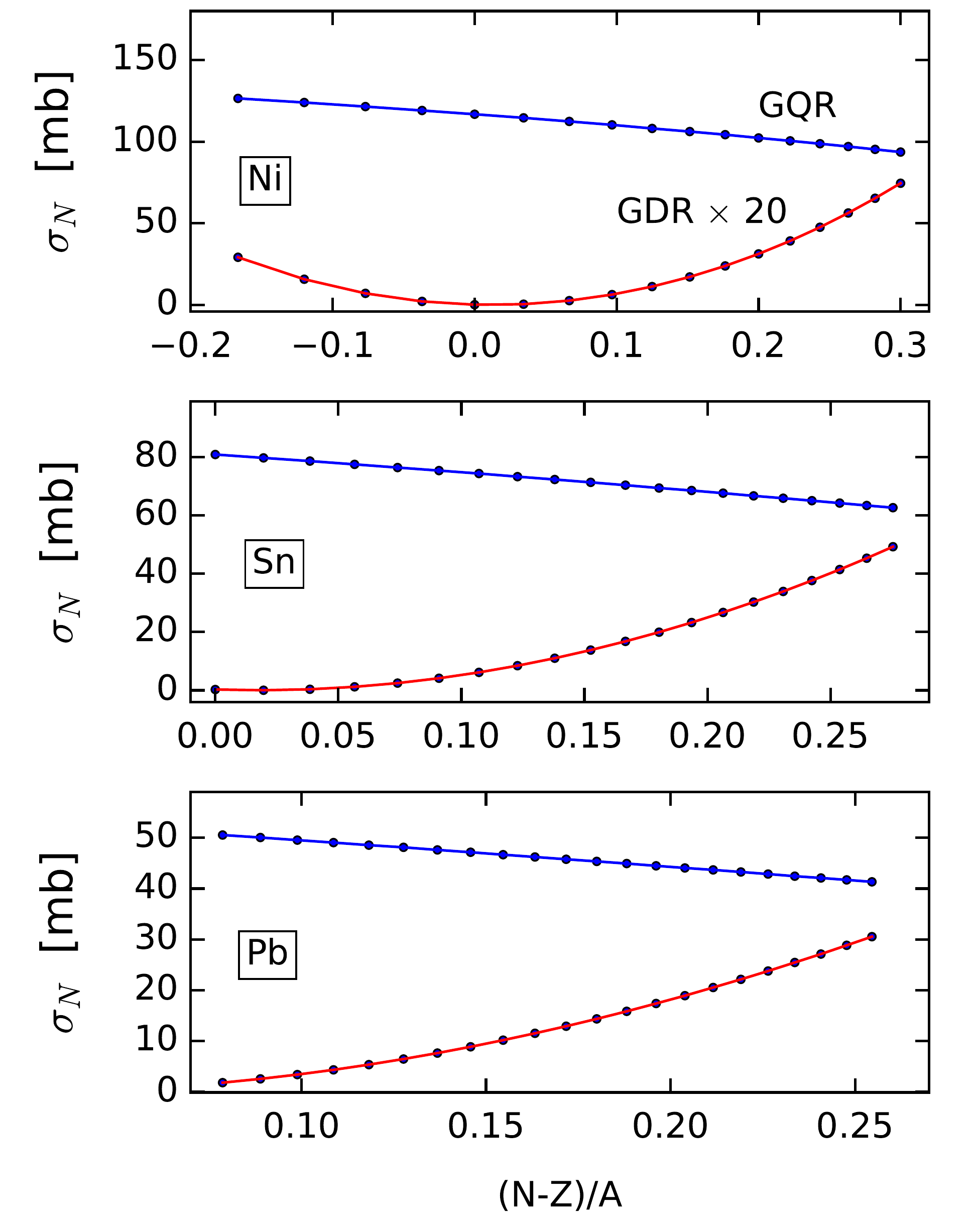}}
\caption{(color online) Cross sections for the nuclear excitation of the ISGQR (GQR) and IVGDR (GDR) in nickel, tin and lead projectiles incident on carbon targets at 1 GeV/nucleon, as a function of the asymmetry coefficient $(N-Z)/A$ of the projectile. The upper curves in each panel are for the excitation of ISGQR and the lower ones for the excitation of IVGDR multiplied by 20 for visualization purposes.}
\label{fig4}
\end{figure}

In Figure \ref{fig4} we plot the cross sections for the nuclear excitation of the ISGQR (GQR) and IVGDR (GDR) in nickel, tin and lead projectiles incident on carbon targets at 1 GeV/nucleon, as a function of the asymmetry coefficient $\delta=(N-Z)/A$ of the projectile. The upper curves in each panel are for the excitation of ISGQR and the lower ones for the excitation of IVGDR multiplied by 20 for visualization purposes. For example, with $^{208}$Pb  projectiles the cross sections are 43.27 mb  and 1.11 mb, for the ISGQR and the IVGDR, respectively.

One notices that the IVGDR cross sections are basically zero for $N=Z$, where one expects a negligible neutron skin. Nickel  exhibits a non-negligible proton skin for light isotopes and as a consequence one sees a reversing trend of the IVGDR excitation cross section around $\delta=0$. However, the cross sections for the IVGDR excitation are at least a factor of 20 smaller than those for the ISGQR and therefore negligible for the purposes of extracting the neutron skin at these bombarding energies. However, this method has been used at lower energies, around and below 100 MeV/nucleon, by using differential cross sections which are able to discern markedly between the $L=1$ and $L=2$ angular distributions and the energy dependence of the cross sections \cite{Kras94}.

The ISGQR cross sections decrease along an isotopic chain as the neutron numbers increase. This can be understood as due to the decrease of the deformation parameter $\beta_2$ with the increase of the ISGQR centroid energy with mass number,  as inferred from Eq. \eqref{beta2}.

There are much larger uncertainties in the theoretical treatment of nuclear excitation in high energy collisions than  the Coulomb excitation case described in the previous section. The Coulomb interaction is well known  while the optical potentials entering the deformed model nuclear interaction of Eq.  \eqref{ul} are not. For the results presented in Fig. \ref{fig4} we have used the ``t-$\rho\rho$'' interaction \cite{BCG03,Ray79}. If instead we use the M3Y interaction \cite{M3Y} with an equal ratio of real to imaginary parts, we get cross sections 50\% smaller, whereas if we use  the JLM interaction \cite{JLM} in the same way, the nuclear excitation cross sections become 30\%-40\% larger. Therefore, at least a factor of 2 uncertainty in the calculations arises due to the nuclear excitation, and there is not much one can do to improve this scenario with the state of the art knowledge of nuclear excitation in high energy collisions. In fact, at relativistic energy collisions such as 1 GeV/nucleon, one needs a four-potential to accommodate Lorentz covariance. For a discussion, see Ref. \cite{Ber05}. The deformed potential model, as well as the Tassie model \cite{Tass56} often used to describe direct nuclear reactions should be considered as rough approximations. The uncertainties arising from nuclear excitation are difficult to quantify unless one would have a much better theoretical description of nuclear excitation followed by neutron emission in high energy collisions. Such a description lacks in the literature.

It is also worthwhile to mention that for proton targets the deformed potential model yields cross sections that are not much different than the ones displayed in Fig. \ref{fig4}. It is difficult to use different targets to change appreciably the nuclear excitation of giant resonances, although a noticeable change of the Coulomb excitation might occur. Sweeping the bombarding energies from 100 MeV/nucleon to 1 GeV/nucleon does not help because the nuclear excitation cross sections remain nearly unchanged. We thus expect that approximately 50 to 100 mb of cross sections, mainly in the one-neutron decay channel, is hard to control systematically without any other information than the angle integrated cross sections.   

The Coulomb and nuclear excitation of giant resonances followed by neutron emission proceeds through the formation of a compound nucleus, which tends to decay nearly isotropically. Therefore, despite the non-negligible magnitude of the excitation cross sections, one can separate fragments arising from excitation followed by decay from those by direct nucleon removal by devising an experiment setup with detection of fragments moving close to the beam direction. As discussed in Ref. \cite{Aum17}, simulations have shown that the nuclear excitation events can be reliably separated using the angular distribution of fragments.

\subsection{Neutron changing and interaction cross sections}

\begin{figure}[t]
\centerline{
\includegraphics[width=0.95\columnwidth]{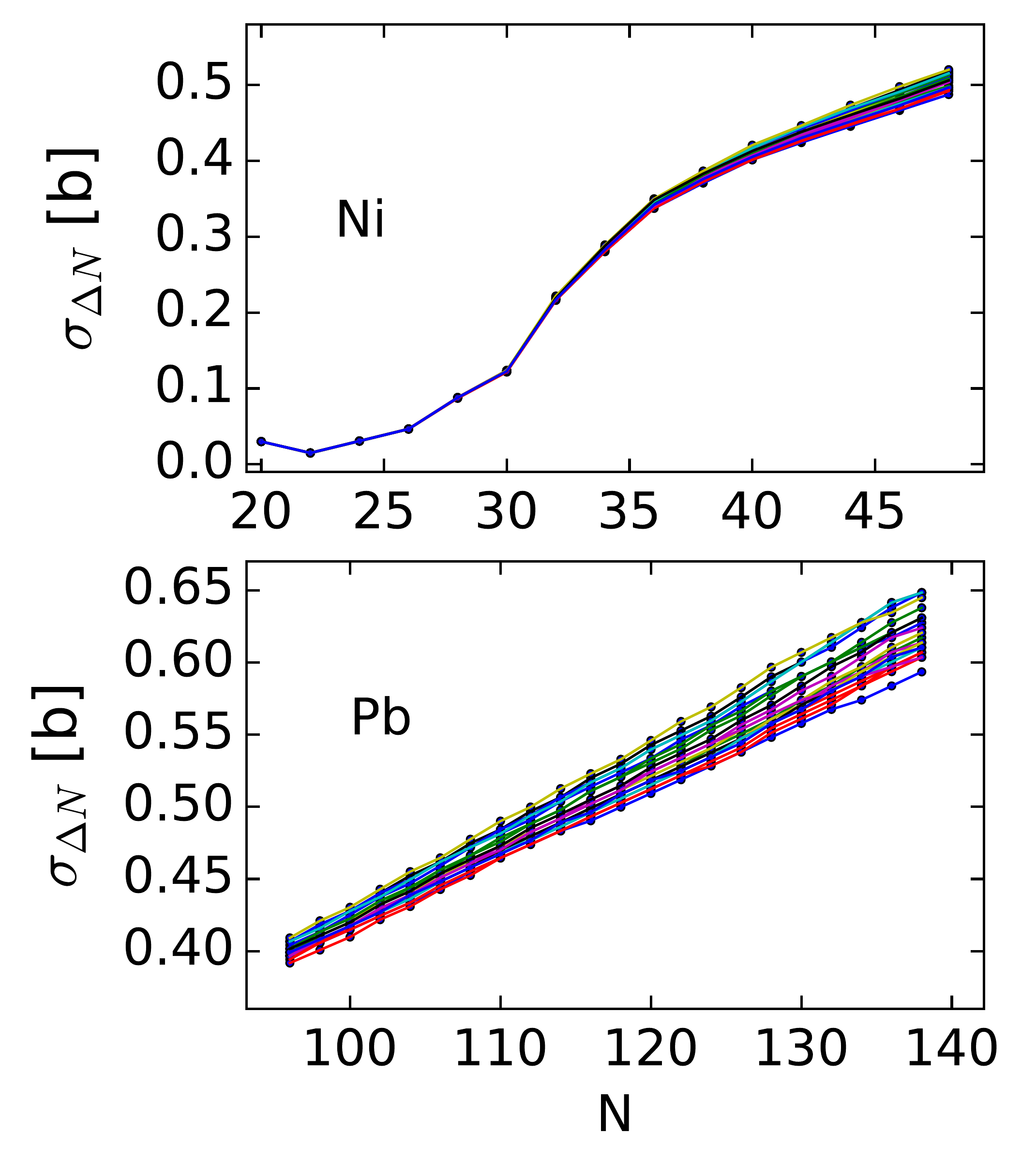}}
\caption{(color online) Neutron-changing cross sections in barns according to Eq. \eqref{sigmaDN}, for nickel (upper panel) and lead (lower panel) isotopes and  the 23 Skyrme interactions mentioned in the text, as a function of the neutron number. The lines are guide to the eyes and each line represents the predictions of one of the Skyrme interactions along an isotopic chain.}
\label{fig5}
\end{figure}

In Figure \ref{fig5} we show the neutron-changing  cross sections in barns, according to Eq. \eqref{sigmaDN}, for nickel (upper panel) and lead (lower panel) isotopes and  the Skyrme interactions adopted, as a function of the neutron number. For nickel we observe a very small dependence on the neutron number with the choice of the Skyrme interaction.  Since nickel is not much larger in size than the carbon target, the nuclear size variation along the isotopic chain with the Skyrme interaction are not large enough to yield a sizable variation of the cross sections. For the heaviest stable nickel isotope, $^{64}$Ni, the cross sections vary within the range $337-350$ mb, which is only a 4\% sensitivity to the choice of the interaction. For a heavy projectile such as lead, the cross sections show a very interesting dependence on the neutron number. First, for every single Skyrme interaction the cross sections display a linear dependence with the neutron number. This is a robust property that can be employed for predictive purposes. Second, for the heaviest stable lead isotope, $^{208}$Pb, the cross sections vary within the range $537-576$ mb, which is nearly a 7\% dispersion with the choice of the interaction. Therefore, it seems that neutron-changing cross sections can constrain the several Skyrme models appreciably. Moreover, the linear relation between $\sigma_{\Delta N}$ and the neutron number is worthwhile to explore in experimental analysis.  

\begin{figure}[t]
\centerline{
\includegraphics[width=0.95\columnwidth]{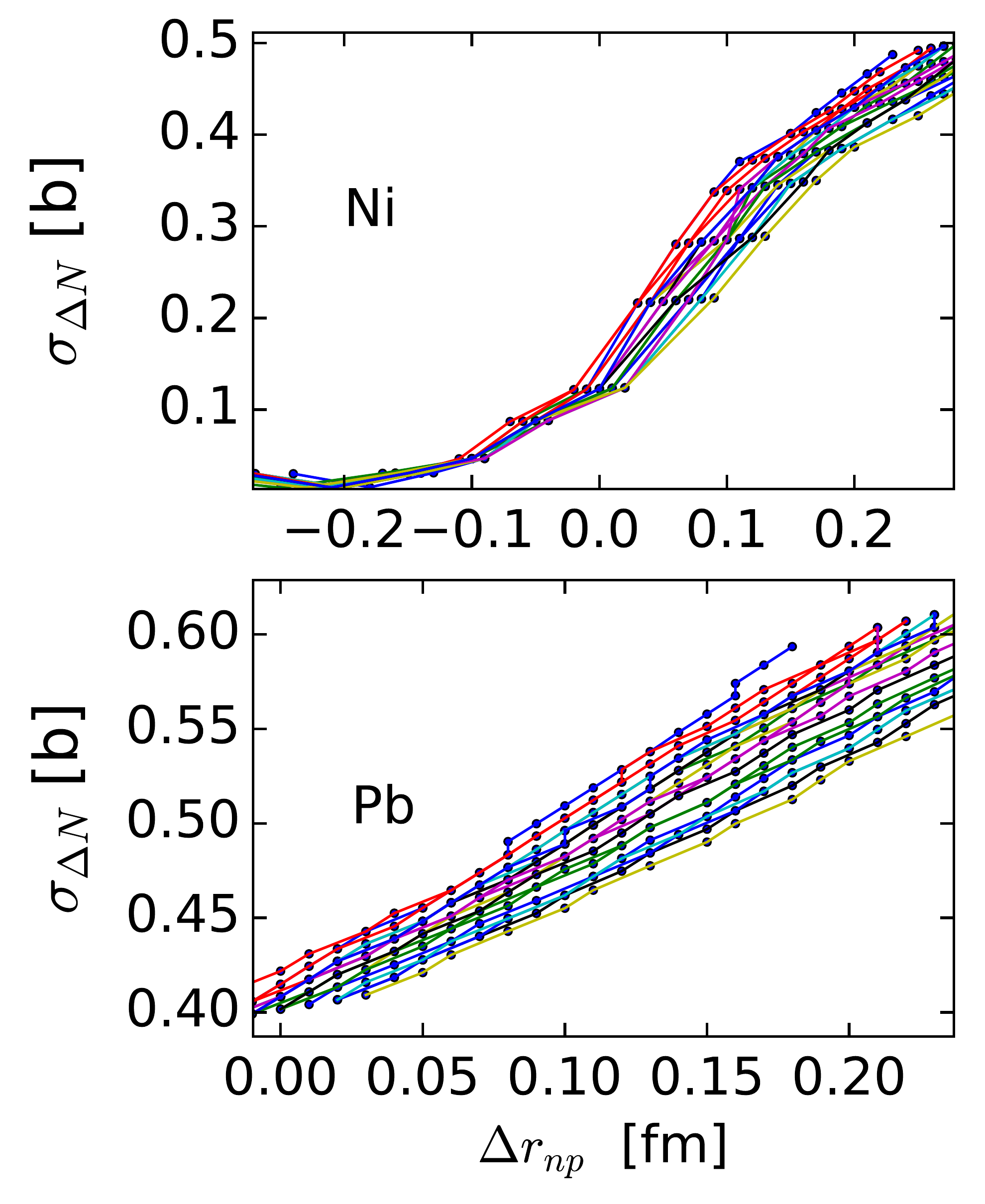}}
\caption{(color online) Same cross sections as in Figure \eqref{fig5} but as a function of the neutron skin $\Delta r_{np}$ in different isotopes and for different Skyrme interactions. The lines are guide to the eyes and represent the predictions of one Skyrme interaction for the neutron skin along an isotopic chain.}
\label{fig6}
\end{figure}

It is justifiable to plot the same data displayed  in Figure \ref{fig5} but as a function of the neutron skin in the different isotopes. This is shown in Figure \ref{fig6}. Now a row of vertical points do not correspond to the same isotope, but each curve along an isotopic chain corresponds to a single Skyrme interaction. Since $\Delta r_{np}$ and the neutron number are correlated (see Figure \ref{fig1}), there is no additional information compared to Figure \ref{fig5}, but one can use this relation to infer the accuracy needed to extract  a given $\Delta r_{np}$. For a neutron skin value of 0.15 fm in Ni and Pb isotopes, the cross sections vary within the range $0.32 - 0.42$ b and $0.47 - 0.52$ b, respectively. These variations correspond to sensitivities of the neutron skin with Skyrme interactions within 20\% for nickel and 10\% for lead isotopes.  Graphs of the sort of Figure \ref{fig6} are only useful if a large number of projectile isotopes are used in an experimental campaign. As we mentioned above, the graph is a combination of  theoretical predictions in Figures \ref{fig1} and \ref{fig5}.

\begin{figure}[t]
\centerline{
\includegraphics[width=0.95\columnwidth]{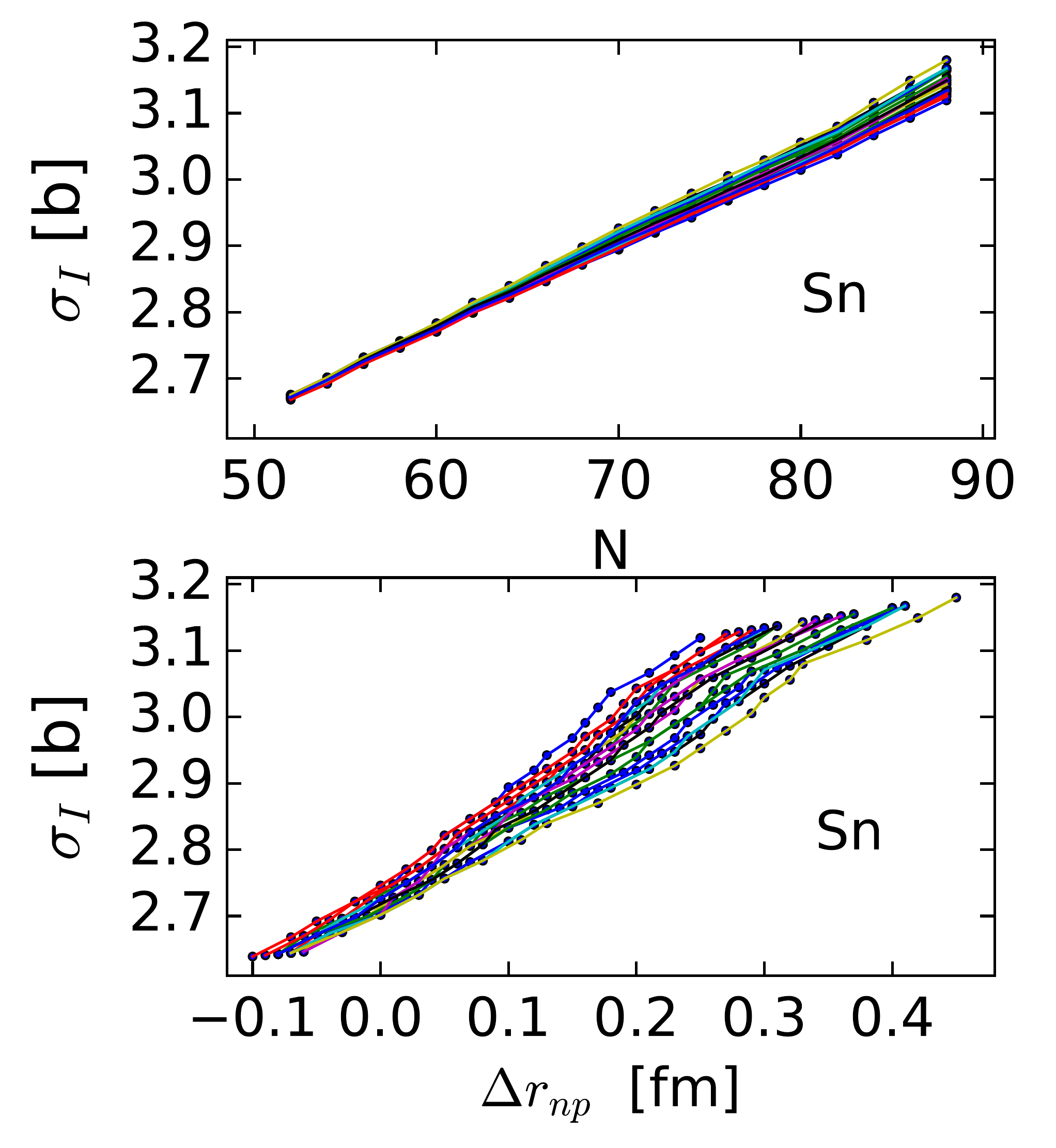}}
\caption{(color online) Total interaction cross sections in barns for tin isotopes incident on carbon targets at 1 GeV/nucleon, according to Eq. \eqref{sigma}, and  several Skyrme interactions. The upper panel shows results as a function of the neutron number $N$, whereas the lower panel displays the same data as a function of the neutron skin $\Delta r_{np}$.}
\label{fig7}
\end{figure}

To emphasize the difference between the cross sections displayed as function of the neutron number and those displayed as function of the neutron skin, we show in Fig. \ref{fig7} the total interaction cross sections in barns for tin isotopes incident on carbon targets at 1 GeV/nucleon. Calculations are done according to Eq. \eqref{sigma}, and for all Skyrme interactions mentioned earlier. The upper panel shows results as a function of the neutron number $N$, whereas the lower panel displays the same data as a function of the neutron skin $\Delta r_{np}$. It is noticeable that the cross sections vary negligibly with the Skyrme interaction for a given isotope. This happens because, for a given isotope, all interactions yield nearly the same total matter density. On the other hand, similar values for neutron skins are obtained for different isotopes with two or more distinct Skyrme interactions. This is evident in the lower panel of Figure \ref{fig7} where we see a much larger variation of $\sigma_{I}$ with $\Delta r_{np}$. Therefore, a combination of measurements of neutron-changing and interaction cross sections can be employed to compare to theoretical microscopic calculations of the nuclear densities. An experimental setup aiming at a 5\% accuracy in these cross sections might be enough to constrain microscopic predictions for the neutron skins.

The major conclusions drawn in this article will not change for different projectile bombarding energies and therefore we do not exploit calculations for different bombarding energies. This feature has been discussed in Ref. \cite{Aum17}. A systematic study of neutron-changing cross sections as a function of the bombarding energy might help in the experimental analysis due to the energy dependence of the nucleon-nucleon (NN) cross section, a crucial input in the calculations entering Eq. \eqref{sigma}. The NN cross section has a pronounced dip at 200 MeV and therefore a mapping of the cross sections for several bombarding energies from 100 MeV/nucleon and up is worth to explore experimentally. 

\begin{figure}[t]
\centerline{
\includegraphics[width=0.95\columnwidth]{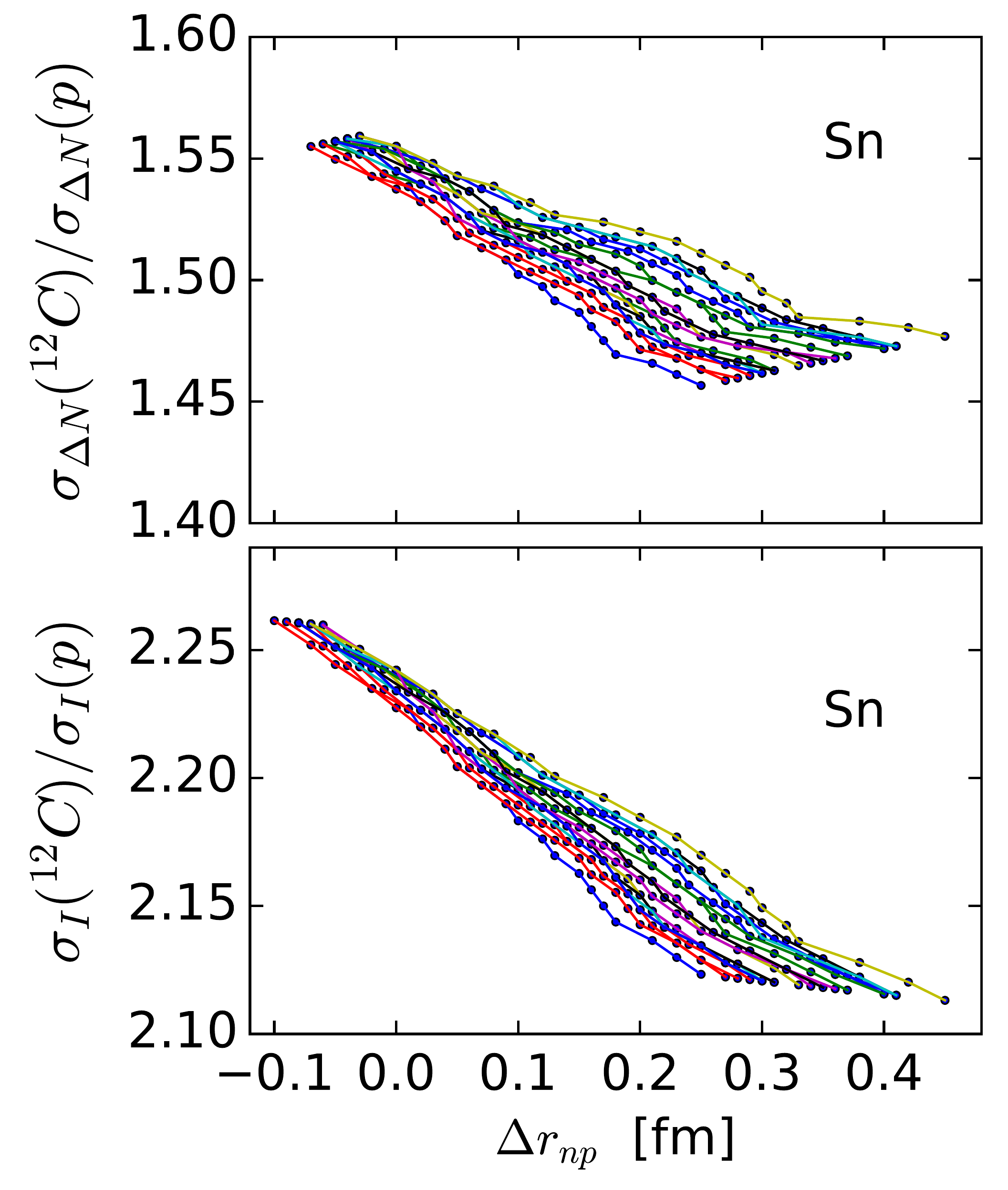}}
\caption{(color online) {\it Upper panel:} Ratio between  the neutron changing cross sections, $\sigma_{\Delta N}$, at 1 GeV/nucleon for tin isotopes obtained with carbon and and with  proton targets,  as a function of the neutron skin, $\Delta r_{np}$. The set of points along a curve correspond to one of the Skyrme interactions used. {\it Lower panel:} The same ratio, but for the total interaction cross sections, $\sigma_{I}$. }
\label{fig8}
\end{figure}

As a final remark, we show in Fig. \ref{fig8} the ratio between  the neutron changing cross sections, $\sigma_{\Delta N}$, at 1 GeV/nucleon for tin isotopes obtained with carbon and and with  proton targets (upper panel) as a function of the neutron skin, $\Delta r_{np}$. In the lower panel we show the same ratio, but for the total interaction cross sections, $\sigma_{I}$. It is clear from the figure that the cross sections obtained with proton targets have a steeper increase with the neutron skin than the cross sections obtained with carbon targets. This is more visible in the ratio of interaction cross sections. Therefore, using carbon and proton targets will allow for a better discrimination of the various Skyrme interactions that could reproduce the experimental data. The measurement of both neutron changing and total interaction cross sections will also help in these studies.

\section{Conclusions}
In summary, in this work we have studied fragmentation reactions as a means to test several microscopic models and their predictions of the neutron-skin thickness in nuclei far from stability. By studying total neutron-changing cross sections one is free from uncertainties of statistical models for compound nucleus decay. On the other hand, Coulomb and nuclear excitation can also influence both total neutron-changing cross sections as well as total interaction cross sections. 

We have shown that the nuclear excitation process followed by neutron emission can add about $50 - 100$ mb to the neutron-changing cross sections.  Only a dedicated experimental setup covering angular distributions of the fragments will be able to eliminate this cross section impurity. We have also shown that Coulomb excitation followed by neutron emission is either negligible or strongly energy dependent, and can be controlled by varying experimental conditions. 

Different microscopic models  will lead to large variations of the neutron skin within an isotopic chain, enough for allowing a discrimination of the best theories to explain the experimental data. This work shows that fragmentation reactions with neutron-rich projectiles at high energies can help us understand the role of the symmetry energy in the equation of state of nuclear matter and its extrapolation to neutron-rich matter needed to explain the structure of neutron stars. 

\section{Acknowledgements}
This work was supported in part by the U.S. DOE grant DE- FG02-08ER41533 and the U.S. National Science Foundation Grant No. 1415656. We thank HIC for FAIR for supporting visits (CAB) to the IKP at the TU-Darmstadt.
 
-----------

\end{document}